\documentclass[12pt, preprint]{aastex}
\usepackage{amsmath}
\usepackage{graphicx}
\usepackage{natbib}

\usepackage[colorlinks=true,citecolor=blue,breaklinks=true,linktocpage=true]{hyperref}
\bibpunct{(}{)}{;}{a}{}{,} 
\usepackage[pagewise]{lineno}
\usepackage{xspace}
\usepackage{hyperref}

\usepackage{multirow}

\def\vec#1{\ensuremath{\mathbf{#1}}}

\newcommand{\vph}{\ensuremath{v_{\rm ph}}}
\newcommand{\vgr}{\ensuremath{v_{\rm gr}}}
\newcommand{\rhoi}{\ensuremath{\rho_{\rm i}}}
\newcommand{\rhoe}{\ensuremath{\rho_{\rm e}}}
\newcommand{\vai}{\ensuremath{v_{\rm Ai}}}
\newcommand{\vae}{\ensuremath{v_{\rm Ae}}}
\newcommand{\vgmin}{\ensuremath{v_{\rm gr}^{\rm min}}}
\newcommand{\omgmin}{\ensuremath{\omega^{\rm min}}}
\newcommand{\vgmax}{\ensuremath{v_{\rm gr}^{\rm max}}}
\newcommand{\omgmax}{\ensuremath{\omega^{\rm max}}}

\usepackage{color}

\shorttitle{Impulsively Generated Sausage Waves in Nonuniform Loops}
\shortauthors{Yu et al.}

\begin{document}


\title{IMPULSIVELY GENERATED SAUSAGE WAVES IN CORONAL TUBES WITH TRANSVERSALLY CONTINUOUS STRUCTURING}

\author{Hui Yu\altaffilmark{1}}
\author{Bo Li\altaffilmark{1}}
    \email{bbl@sdu.edu.cn}
\author{Shao-Xia Chen\altaffilmark{1}}
\author{Ming Xiong\altaffilmark{2}}
\and
\author{Ming-Zhe Guo\altaffilmark{1}}

\altaffiltext{1}{Shandong Provincial Key Laboratory of Optical Astronomy and Solar-Terrestrial Environment,
     Institute of Space Sciences, Shandong University, Weihai 264209, China}
\altaffiltext{2}{State Key Laboratory of Space Weather, National Space Science Center,
Chinese Academy of Sciences, Beijing 100190, China}

\begin{abstract}
The frequency dependence of the longitudinal group speeds of trapped sausage waves plays an important role
    in determining impulsively generated wave trains,
    which have often been invoked to account for
    quasi-periodic signals in coronal loops.
We examine how the group speeds ($\vgr$) depend on angular frequency ($\omega$)
    for sausage modes in pressureless coronal tubes with continuous
    transverse density distributions by solving the dispersion relation pertinent to the case where
    the density inhomogeneity of arbitrary form takes place in a transition layer of arbitrary thickness.
We find that in addition to the transverse lengthscale $l$ and density contrast $\rho_{\rm i}/\rho_{\rm e}$,
    the group speed behavior depends also on the detailed form of the density inhomogeneity.
For parabolic profiles, $\vgr$ always decreases with $\omega$ first before increasing again,
    as happens for the much studied top-hat profiles.
For linear profiles, however, the behavior of the $\omega-\vgr$ curves is more complex.
When $\rho_{\rm i}/\rho_{\rm e} \lesssim 6$, the curves {become} monotonical for large values of $l$.
On the other hand, for higher density contrasts, a local maximum $\vgmax$ exists
    in addition to a local minimum $\vgmin$ when coronal tubes are diffuse.
With time-dependent computations, we show that the different behavior of group speed curves,
    the characteristic speeds $\vgmin$ and $\vgmax$ in particular,
    is reflected in the temporal evolution and Morlet spectra of impulsively generated wave trains.
We conclude that the observed quasi-periodic wave trains not only can be employed to probe
    such key parameters as density contrasts and profile steepness,
    but also have the potential to discriminate between the unknown forms
    of the transverse density distribution.
\end{abstract}

\section{INTRODUCTION}
\label{sec_intro}

The past two decades have seen rapid progress in
    the field of solar magneto-seismology
    \citep[SMS, for recent reviews, see e.g.,][]{2005LRSP....2....3N,2007SoPh..246....3B,2012RSPTA.370.3193D,2016GMS...216..395W,2016SSRv..200...75N}.
Among the rich variety of the low-frequency waves observed in the Sun's atmosphere,
    flare-related quasi-periodic fast propagating (QFP) wave trains have received much attention
    \citep[see][for a recent review]{2014SoPh..289.3233L}.
Their quasi-periods usually ranging from $25$ to $400$ secs,
    these wave trains were
    discovered~\citep{2011ApJ...736L..13L}
    and extensively observed in images acquired with
    the Atmospheric Imaging Assembly on board the Solar Dynamics Observatory
    {
    (SDO/AIA, \citeauthor{2012ApJ...753...52L}~\citeyear{2012ApJ...753...52L};
    \citeauthor{2013A&A...554A.144Y}~\citeyear{2013A&A...554A.144Y};
    \citeauthor{2013SoPh..288..585S}~\citeyear{2013SoPh..288..585S};
    \citeauthor{2014A&A...569A..12N}~\citeyear{2014A&A...569A..12N};
    also see 
    \citeauthor{2012SoPh..275...17L}~\citeyear{2012SoPh..275...17L}
    for the description of the instrument). }
On the other hand, quasi-periodic signals in coronal emissions presumably from density-enhanced loops
    have been known since the 1960's
    (e.g., \citeauthor{1969ApJ...155L.117P}~\citeyear{1969ApJ...155L.117P},
    \citeauthor{1969ApJ...158L.159F}~\citeyear{1969ApJ...158L.159F},
    \citeauthor{1970A&A.....9..159R}~\citeyear{1970A&A.....9..159R},
    \citeauthor{1973SoPh...32..485M}~\citeyear{1973SoPh...32..485M};
    see Table 1 of
    \citeauthor{1999ApJ...520..880A}~\citeyear{1999ApJ...520..880A}
    for a comprehensive compilation of measurements prior to 2000).
The quasi-periods $P$ of a considerable fraction of these signals
    were of the order of seconds.
While these measurements were largely spatially unresolved,
    more recent high-cadence instruments imaging the corona at total eclipses
    also indicated the presence in coronal loops
    of quasi-periodic signals both with $P\sim 4-7$ seconds~\citep{2001MNRAS.326..428W,2002MNRAS.336..747W,2003A&A...406..709K}
    and with $P\sim 6-25$ seconds~\citep{2016SoPh..291..155S}.
In addition, quasi-periodic pulsations (QPPs) in the lightcurves of solar flares with similar periods
    have also been measured using imaging instruments such as
    the Nobeyama Radioheliograph \citep[NoRH, e.g.,][]{2001ApJ...562L.103A,  2003A&A...412L...7N, 2005A&A...439..727M, 2013SoPh..284..559K},
    SDO/AIA \citep[e.g.,][]{2012ApJ...755..113S},
    and more recently with the Interface Region Imaging Spectrograph
    {
    (IRIS, \citeauthor{2016ApJ...823L..16T}~\citeyear{2016ApJ...823L..16T};
    see also \citeauthor{2014SoPh..289.2733D}~\citeyear{2014SoPh..289.2733D} for the description
    of IRIS).
    }

Interestingly, quasi-periodic waves are not necessarily connected to
    quasi-periodic activities of the sources but can also be attributed
    to their impulsive generation
    {(see \citeauthor{2015ApJ...809L..17J}~\citeyear{2015ApJ...809L..17J}
    and \citeauthor{2015ApJ...815L..16S}~\citeyear{2015ApJ...815L..16S} for recent observational evidence,
    \citeauthor{2015ApJ...800..111Y}~\citeyear{2015ApJ...800..111Y}
    and \citeauthor{2016ApJS..224...30Y}~\citeyear{2016ApJS..224...30Y}
    for numerical demonstrations).
As explained by~\citet{1983Natur.305..688R, 1984ApJ...279..857R},}
    the key here is wave dispersion.
For simplicity, consider fast sausage waves in a density-enhanced coronal loop
    where the transverse density distribution is in a top-hat fashion,
    characterized by the internal ($\rhoi$) and external values ($\rhoe$).
Let $\vai$ and $\vae$ denote the internal and external Alfv\'en speeds, respectively
    ($\vae > \vai$).
It is well-known that two regimes exist for sausage waves, depending on how
    the axial wavenumber $k$ compares with
    some critical value $k_{\rm c}$.
The leaky regime arises when $k < k_{\rm c}$,
    whereby waves lose their energy by radiating fast waves into the surroundings
    \citep{1978SoPh...58..165M,1982SoPh...75....3S,1986SoPh..103..277C}.
When $k > k_{\rm c}$, however, the trapped regime results and
    wave energy is well confined to coronal loops.
Trapped sausage waves are substantially dispersive such that
    when the angular frequency $\omega$ increases from $\omega_{\rm c} = k_{\rm c} \vae$,
    the axial group speed $\vgr$
    first decreases to a local minimum $\vgr^{\rm min}$ at $\omega^{\rm min}$,
    before increasing towards $\vai$.
Graphically speaking, this means that the $\omega-\vgr$ curve comprises two portions:
    in one portion a horizontal line representing a constant $\omega$ intersects the curve
    at a single point ($v_{\rm Ae} > \vgr > v_{\rm Ai}$),
    whereas multiple intersections exist in the other ($v_{\rm Ai} > \vgr > \vgmin$).
Let the first (second) portion be labeled ``S'' (``M''), short for ``single'' (``multiple'').
\citet{1984ApJ...279..857R} predicted that in response to an impulsive internal axisymmetric source,
    the signal in the loop at a distance $h$ sufficiently far away
    comprises three phases
    \citep[see also][for a heuristic discussion]{1986NASCP2449..347E}.
For the time interval $h/v_{\rm Ae} < t < h/v_{\rm Ai}$,
    wavepackets in portion S are relevant and individual wavepackets
    with progressively low group speeds arrive consecutively.
Portion M then becomes relevant when $h/v_{\rm Ai} < t < h/\vgmin$,
    whereby multiple wavepackets with the same group speed but different frequencies
    arrive simultaneously.
The signal now is expected to be stronger than in the first phase
    due to the superposition of multiple wavepackets.
When $t> h/\vgmin$, no incoming wavepackets are expected, resulting in some decaying signal
    oscillating at an angular frequency $\omega^{\rm min}$.
These three phases are traditionally termed the periodic, quasi-periodic and decay (or Airy) phases.

The theoretically expected evolution of impulsively generated sausage waves
   has been shown to be robust by both analytical and numerical studies.
Examining the response of a transversally discontinuous
   coronal tube to localized, axisymmetric, footpoint motions,
   \citet{1996ApJ...472..398B} showed analytically that
   an alternative interpretation of the periodic phase
   is that wavefronts that initially leak out of the tube
   will return back in view of the continuity of the transverse Lagrangian displacement.
Also examining coronal tubes with top-hat transverse density profiles,
   the more recent analytical study by \citet{2015ApJ...806...56O}
   showed that impulsively generated waves can be examined in terms of the partition of
   the energy contained in the initial perturbation between proper and improper modes.
On the other hand, impulsively generated sausage waves
   have also been numerically examined in both coronal slabs
   \citep{1993SoPh..144..101M, 1994SoPh..151..305M, 2004MNRAS.349..705N, 2013A&A...560A..97P, 2016ApJ...826...78Y}
   and coronal tubes \citep{2007A&A...462.1127S, 2015ApJ...814..135S}, for which
   the density is transversally distributed in a continuous manner.
What these studies suggest is that, while such factors as the axial extent of the initial perturbation
   \citep{2015ApJ...806...56O} and density profile steepness \citep{2004MNRAS.349..705N,2015ApJ...814..135S}
   are important,
   impulsively generated wave trains
   can still to a large extent be understood in the framework proposed by \citet{1984ApJ...279..857R}.
In particular, the numerical studies by \citet{2004MNRAS.349..705N} and \citet{2015ApJ...814..135S}
   showed that the period and amplitude modulations in the wave trains transform into
   tadpole-shaped Morlet spectra.
This spectral feature was actually seen in both radio \citep[e.g.,][]{2010ITPS...38.2243J, 2013A&A...550A...1K}
   and optical measurements \citep[e.g.,][]{2003A&A...406..709K, 2016SoPh..291..155S}.

The temporal and wavelet features of impulsively generated waves
   can help yield
   such key information as the internal Alfv\'en speed, density contrast between loops and their surroundings,
   as well as the location of the source
   \citep[e.g.,][]{1984ApJ...279..857R, 2008IAUS..247....3R}.
However, such applications are primarily based on the theoretical results for fast sausage waves
   in coronal tubes with a top-hat transverse density profile~\citep{1983SoPh...88..179E}.
While theoretical results are known
   for a limited number of continuous density profiles,
   they are either not fully developed \citep{1988A&A...192..343E}
   or have not been applied to the context of impulsively generated waves
   \citep{2014A&A...572A..60L,2015ApJ...810...87L}.
In an effort of developing a more general theory for sausage waves,
   we worked in the framework of cold MHD and examined coronal tubes with transverse density profiles
   that comprise a uniform cord, an external uniform medium, and a transition layer (TL) connecting the two
   \citep[][hereafter Paper I]{2015ApJ...812...22C}.
The TL can be of arbitrary width and the density profile therein can be of arbitrary form.
We further showed that it is straightforward to eliminate the requirement for the density profile to involve
   a uniform cord \citep{2016SoPh..291..877G}.
The present manuscript intends to make a fuller use of the theory developed in Paper I,
   and to examine how different prescriptions for the transverse density profiles
   influence the properties of impulsively generated sausage waves.

This manuscript is organized as follows.
Section~\ref{sec_tophat_para} extends some previous theoretical studies
    on sausage waves in coronal tubes with two simple density profiles.
The purpose is to offer some new analytical results such that the
    the frequency dependence of the group speeds can be better understood.
Section~\ref{sec_series} builds on our Paper I and provides a detailed examination on
    how different profile choices impact the group speed curves.
The consequences on the temporal evolution of impulsively generated waves are also discussed therein
    with the aid of time-dependent computations.
Finally, Section~\ref{sec_conc} closes this manuscript with our summary
    and some concluding remarks.

\section{SAUSAGE WAVES IN CORONAL LOOPS WITH TWO SIMPLE DENSITY PROFILES}
\label{sec_tophat_para}

{In view of applications to the solar corona where gas pressure is usually much smaller
    than magnetic pressure},
    we adopt the cold MHD description {in which
    the gas to magnetic pressure ratio is taken to be zero}.
We model coronal loops as straight magnetic tubes with a mean radius $R$
    and directed in the $z$-direction in a cylindrical coordinate system $(r, \theta, z)$.
The magnetic field is uniform and also in the $z$-direction, $\vec{B} = B\hat{z}$.
We proceed with a standard eigenmode analysis by assuming that all perturbations are
    proportional to $\exp\left(i k z - i\omega t\right)$.
Note that both $\omega$ and $k$ are taken to be real,
    given that only trapped modes are of interest here.
In addition, only the lowest-order sausage waves will be examined {throughout}.

{As stated in the introduction, this study will be focused on the effects of different transverse density profiles
    on impulsively generated sausage waves.
However, it is evidently impossible to exhaust all possible choices for the largely unknown
    form of density distribution.
We therefore choose to examine a set of profiles that are often invoked in examinations
    of kink modes~\citep[e.g.,][]{2002ApJ...577..475R, 2004ApJ...606.1223V, 2014ApJ...781..111S}.
Figure~\ref{fig_illus_profile} displays the chosen profiles where for illustration purposes,
    a value of $10$ is adopted for the density contrast
    $\rho_{\rm i}/\rho_{\rm e}$, the ratio of the density at the loop axis to that far from the loop.
The profiles in Figure~\ref{fig_illus_profile}b pertains to Section~\ref{sec_series} and will be explained therein.
This section will examine the two profiles in Figure~\ref{fig_illus_profile}a where they are labeled ``top-hat'' and ``fully parabolic''.
As will be shown, compact closed-form expressions can be found for sausage waves in coronal loops with these two simple profiles.
}

\subsection{Top-hat Profiles}
\label{sec_sub_tophat}
Let us start with a top-hat transverse density profile
    {(illustrated by the black curve in Figure~\ref{fig_illus_profile}a)},
\begin{eqnarray}
 {\rho}(r)=\left\{
   \begin{array}{ll}
   \rhoi,    & 0 \le r < R \\
   \rhoe,    & r > R ,
   \end{array}
   \right.
\label{eq_rho_tophat}
\end{eqnarray}
    where the subscripts ${\rm i}$ and ${\rm e}$
    denote the internal and external values, respectively.
The dispersion relation (DR) for trapped sausage waves reads~\citep[e.g.,][]{1983SoPh...88..179E}
\begin{eqnarray}
    n \displaystyle\frac{J_0(n R)}{J_1(n R)}
 =- m \displaystyle\frac{K_0(m R)}{K_1(m R)} ,
\label{eq_DR_tophat}
\end{eqnarray}
    where $J_j$ ($K_j$) represents Bessel function of the first kind
    (modified Bessel function of the second kind) with $j=0, 1$.
In addition,
\begin{eqnarray}
  n^2 = \displaystyle\frac{\omega^2}{\vai^2} - k^2 >0, \hspace{0.5cm}
  m^2 = k^2 - \displaystyle\frac{\omega^2}{\vae^2} >0,
\label{eq_def_nm}
\end{eqnarray}
    in which $v_{\rm Ai,e} = B/\sqrt{4\pi \rho_{\rm i,e}}$ is the Alfv\'en speed.
To proceed, let $\vph \equiv \omega/k$ and $\vgr = {\rm d}\omega/{\rm d} k$
    denote the axial phase and group
    speeds, respectively.

Some analytical results for large $kR$
   can be readily found.
It turns out that~\citep{2014A&A...568A..31L,2016RAA....16f...7Y}
\begin{eqnarray}
   \displaystyle\frac{\vph}{\vai} \approx \sqrt{1+\frac{j_{1,l}^2}{(kR)^2}} ,
\label{eq_vp_bigK_tophat}
\end{eqnarray}
   where $j_{1,l}$ denotes the $l$-th zero of $J_1$ with $j_{1,0} = 3.8317$.
Consequently, $\vgr$ at large $kR$ can be approximated by
\begin{eqnarray}
   \displaystyle\frac{\vgr}{\vai} \approx \frac{1}{\sqrt{1+{j_{1,l}^2}/{(kR)^2}}} .
\label{eq_vg_bigK_tophat}
\end{eqnarray}
This suggests that as $kR$ approaches infinity,
   $\vph$ ($\vgr$) gradually approaches $\vai$
   from above (below).

\subsection{Fully Parabolic Profiles}
\label{sec_sub_para}
Now consider a continuous density profile given by
\begin{eqnarray}
 {\rho}(r)=\left\{
   \begin{array}{ll}
   \rhoi\left[1-\left(1-\displaystyle\frac{\rhoe}{\rhoi}\right)
        \left(\displaystyle\frac{r}{r_{\rm e}}\right)^2\right],
        & 0 \le r \leq r_{\rm e} \\
   \rhoe,    & r \ge r_{\rm e} ,
   \end{array}
   \right.
\label{eq_rho_fullypara}
\end{eqnarray}
    where $r_{\rm e} = 2R$ denotes the loop-external-medium interface,
    {with $R$ being the mean tube radius}.
For future reference, let this profile be denoted by ``fully parabolic'',
    {which is illustrated by the red curve in Figure~\ref{fig_illus_profile}a}.
A similar profile was considered by \citet{1965PhFl....8..507P},
    where the external medium was taken to be vacuum ($\rhoe =0$).
Omitting the details, we extend the mathematical procedure therein
    to account for a finite $\rhoe$, with the resulting DR reading
\begin{eqnarray}
   - (m r_{\rm e}) \displaystyle\frac{K_0(m r_{\rm e})}{K_1(m r_{\rm e})}
 = 2 - p + \alpha p \displaystyle\frac{M(\alpha+1, 3, p)}{M(\alpha, 2, p)} ,
\label{eq_DR_para}
\end{eqnarray}
   where
\begin{eqnarray*}
   && p = \displaystyle\frac{\omega r_{\rm e}}{\vai}
      \sqrt{1-\displaystyle\frac{\rhoe}{\rhoi}} ,  \\
   && \alpha = 1-\displaystyle\frac{\left(\omega r_{\rm e}/\vai\right)^2-\left(k r_{\rm e}\right)^2}{4 p} .
\end{eqnarray*}
In addition,
\begin{eqnarray*}
     M(a, b, x) = 1+\frac{a}{b}x+\frac{a(a+1)}{b(b+1)}\frac{x^2}{2!} + \cdots
\end{eqnarray*}
   is Kummer's function.
When deriving Equation~(\ref{eq_DR_para}), we have used the fact that
\begin{eqnarray*}
 \displaystyle\frac{{\rm d}M(a, b, x)}{{\rm d}x} = \displaystyle\frac{a}{b}M(a+1, b+1, x).
\end{eqnarray*}

Approximate expressions for $\vph$ and $\vgr$ at large $kR$ can also be found by generalizing the results
    of \citet{1965PhFl....8..507P} where $\rhoe = 0$.
%
For the lowest order sausage waves, we find that $\alpha$ rapidly approaches zero with increasing $kR$.
Consequently,
\begin{eqnarray}
  \displaystyle\frac{\vph}{\vai}
  \approx 1+\displaystyle\frac{\sqrt{4(1-\rhoe/\rhoi)}}{k r_{\rm e}}
                +\displaystyle\frac{2(1-\rhoe/\rhoi)}{(k r_{\rm e})^2} ,
  \label{eq_vph_bigK_para}
\end{eqnarray}
   and
\begin{eqnarray}
  \displaystyle\frac{\vgr}{\vai}
  \approx 1-\displaystyle\frac{2(1-\rhoe/\rhoi)}{(k r_{\rm e})^2} .
  \label{eq_vg_bigK_para}
\end{eqnarray}
As in the top-hat case, one expects to see that
   $\vph$ ($\vgr$) approaches $\vai$
   from above (below).

\subsection{Group Speed Curves}
\label{sec_sub_tophatpara_disp_diag}
Figure~\ref{fig_vphvgkdep_tophatfullypara} presents the $kR$ dependence of
   the phase {(the upper row)} and group {(lower)} speeds
   for both a top-hat (the black curves) and a fully parabolic (red) profile.
For illustration purposes,
   {two values, $3$ (the left column) and $10$ (right), are adopted for the density ratio $\rhoi/\rhoe$}.
In addition to the full numerical solutions to the DRs (the solid lines), we also plot
   their approximate expressions for large axial wavenumbers (dashed).
{From Figures~\ref{fig_vphvgkdep_tophatfullypara}a and \ref{fig_vphvgkdep_tophatfullypara}c}
   one sees that for both profiles,
   trapped waves exist only when $k$ exceeds some critical value $k_{\rm c}$,
   beyond which $\vph$ decreases monotonically with $kR$ and
   approaches $\vai$ as expected.
{In addition, one sees that for both density contrasts, the dimensionless critical wavenumber $k_{\rm c}R$
   is smaller for the fully parabolic than for the top-hat profile.
Actually this is true for $\rhoi/\rhoe$ in an extensive range between $2$ and $1000$.
}
As for the group speeds, one sees from {Figures~\ref{fig_vphvgkdep_tophatfullypara}b and  \ref{fig_vphvgkdep_tophatfullypara}d}
   that $\vgr$ decreases sharply first with $kR$
   before increasing towards $\vai$ for large $kR$.
The asymptotic $k$ dependence of both $\vph$ and $\vgr$ {for the fully parabolic profile
   is well approximated by the analytical expressions (\ref{eq_vph_bigK_para}) and (\ref{eq_vg_bigK_para}).
For instance, when $\rho_{i}/\rho_{e} =10$, the red dashed lines can hardly be told apart from
    the solid ones when $kR \gtrsim 3$.
For the top-hat profiles, however, one sees that the black dashed curves
   converge to the solid ones
   only for sufficiently large $kR$.
When $\rho_{i}/\rho_{e} =10$,
   this happens for $kR\gtrsim 20$.
The reason for the different performance of the approximate expressions at large $kR$
   is that the density contrast does not appear in Equations~(\ref{eq_vp_bigK_tophat}) and (\ref{eq_vg_bigK_tophat})
   for top-hat profiles, whereas it is explicitly involved for fully parabolic profiles
   (see Equations~\ref{eq_vph_bigK_para} and \ref{eq_vg_bigK_para}).
   }

The wavenumber dependence is readily translated into frequency dependence, which is displayed
    in Figure~\ref{fig_vgr_tophatfullypara}.
{Once again two different density contrasts, $3$ (the dash-dotted curves) and $10$ (solid)
    are shown for illustration purposes.
   }
One sees the typical behavior for $\vgr$ to posses a local minimum $\vgmin$.
This behavior is well-known for top-hat profiles \citep{1983Natur.305..688R, 1984ApJ...279..857R},
   and from Equation~(\ref{eq_vg_bigK_tophat})
   we know that
   {this $\omega$ dependence of the group speed curves holds regardless of
   the density contrast, even though only two values are examined here.}
However, the group speed curves for fully parabolic profiles
   are qualitatively different from the only relevant study
   where $\vgr$ was shown to increase monotonically  with $\omega$ from zero to $\vai$
   when $\rho_{\rm e} =0$
   \citep[][Figure~2]{1965PhFl....8..507P}.
In this regard, the red {curves in Figure~\ref{fig_vgr_tophatfullypara} show} that a finite $\rho_{\rm e}$, or physically speaking the finite elasticity
   of the external medium, is crucial for $\vgr$ to posses a local minimum
   \citep[see also the discussion in][Section 3.2]{1988A&A...192..343E}.
In addition, the asymptotic expression (\ref{eq_vg_bigK_para}) indicates that
   such a local minimum exists for arbitrary {$\rho_{\rm e}$, as long as it is finite}.

\section{SAUSAGE WAVES IN CORONAL TUBES WITH ARBITRARY DENSITY PROFILES}
\label{sec_series}

At this point, we note that both top-hat and fully parabolic profiles
   lead to the classical group speed curves possessing a local minimum,
   and this behavior holds for arbitrary density contrasts as long as
   the external density does not vanish.
Let $+$ ($-$) denote the tendency for $\vgr$ to increase (decrease) with $\omega$.
Then typically one sees this $-/+$ behavior.
However, will other possibilities appear for some other density profiles?
This is addressed in the present section.

\subsection{Profile Description and Dispersion Relation}

We choose to examine some density profiles adopted in Paper I, where
    the transverse density distribution comprises a transition layer (TL) of width $l$
    connecting a uniform cord and a uniform external medium,
\begin{eqnarray}
 {\rho}(r)=\left\{
   \begin{array}{ll}
   \rho_{\rm i},    & 0\le r \leq r_{\rm i} = R-l/2, \\
   \rho_{\rm tr}(r),& r_{\rm i} \le r \le r_{\rm e} = R+l/2,\\
   \rho_{\rm e},    & r \ge r_{\rm e} .
   \end{array}
   \right.
\label{eq_rho_profile}
\end{eqnarray}
The profile between $[r_{\rm i}, r_{\rm e}]$ is such that
   the equilibrium density $\rho$ decreases continuously from
   the internal value $\rho_{\rm i}$
   to the external one $\rho_{\rm e}$.
Evidently, the thickness of this TL ($l$),
   {a measure of profile steepness,
   is bounded by $0$ and $2 R$.
In this sense, the dimensionless parameter $l/2R$ can be used to
   distinguish between thin (with, say, $l/2R \le 0.5$)
   and thick (with $l/2R \ge 0.5$) transition layers.
   }
Two prescriptions will be explored here,
\begin{eqnarray}
\label{eq_rho_tr}
   \rho_{\rm tr}(r)=\left\{
   \begin{array}{ll}
   \rho_{\rm i}-\displaystyle\frac{\rho_{\rm i}-\rho_{\rm e}}{l}\left(r-r_{\rm i}\right),   & {\rm linear~{TL}},
   \\[0.3cm]
   \rho_{\rm i}-\displaystyle\frac{\rho_{\rm i}-\rho_{\rm e}}{l^2}\left(r-r_{\rm i}\right)^2,& {\rm parabolic~{TL}} .\\[0.3cm]
   \end{array}
   \right.
\end{eqnarray}
{For an illustration of these two profiles, please see Figure~\ref{fig_illus_profile}b
   where $\rho_{\rm i}/\rho_{\rm e}$ and $l$ are arbitrarily chosen to be $10$ and $R$, respectively.}
We note that the profile labeled ``parabolic {TL}''
   recovers the fully parabolic profile (\ref{eq_rho_fullypara}) when $l$
   approaches $2R$.

The DR for sausage waves for arbitrary prescriptions of $\rho_{\rm tr}(r)$
    in a TL of arbitrary thickness was given by Equation~(17) in Paper I.
Focusing on trapped waves, one can readily reformulate this DR as
\begin{eqnarray}
 \displaystyle
 \frac{\tilde{\xi}_{1}(x_{\rm i})+r_{\rm i}\tilde{\xi}'_{1}(x_{\rm i})
     -\displaystyle\frac{n r_{\rm i}J_0(n r_{\rm i})}{J_1(n r_{\rm i})}\tilde{\xi}_{1}(x_{\rm i})}
      {\tilde{\xi}_{2}(x_{\rm i})+r_{\rm i}\tilde{\xi}'_{2}(x_{\rm i})
     -\displaystyle\frac{n r_{\rm i}J_0(n r_{\rm i})}{J_1(n r_{\rm i})}\tilde{\xi}_{2}(x_{\rm i})}
=\displaystyle
 \frac{\tilde{\xi}_{1}(x_{\rm e})+r_{\rm e}\tilde{\xi}'_{1}(x_{\rm e})
     -\displaystyle\frac{m r_{\rm e}K_0(m r_{\rm e})}{K_1(m r_{\rm e})}\tilde{\xi}_{1}(x_{\rm e})}
      {\tilde{\xi}_{2}(x_{\rm e})+r_{\rm e}\tilde{\xi}'_{2}(x_{\rm e})
     -\displaystyle\frac{m r_{\rm e}K_0(m r_{\rm e})}{K_1(m r_{\rm e})}\tilde{\xi}_{2}(x_{\rm e})}~
\label{eq_DR_arbi}
\end{eqnarray}
Here $m$ and $n$ have already been defined by Equation~(\ref{eq_def_nm}).
In addition, $x \equiv r-R$ and $x_{\rm i, e} = r_{\rm i, e} -R = \mp l/2$.
Furthermore, $\tilde{\xi}_{1}$ and $\tilde{\xi}_{2}$ denote two linearly independent solutions for
    the Lagrangian displacement in the TL.
Both are expressed as a regular series expansion around $x=0$,
\begin{eqnarray*}
   \tilde{\xi}_{1}(x)=\sum^\infty_{n=0}a_n x^{n}~~\mbox{and}~~
   \tilde{\xi}_{2}(x)=\sum^\infty_{n=0}b_n x^{n} .
\end{eqnarray*}
Without loss of generality, we choose
\begin{eqnarray*}
   a_0=R,~~a_1=0,~~b_0=0,~~b_1 =1 .
\end{eqnarray*}
The rest of the coefficients $a_n$ and $b_n$ are given by Equation~(11) in paper I,
    and contain the information on the density distribution.
Furthermore, the prime $'$ denotes the derivative of $\tilde{\xi}_{1, 2}$ with respect to $x$.

\subsection{Group Speed Curves}
Let us first examine parabolic {TL} profiles, which
    approach top-hat profiles when $l \rightarrow 0$,
    and approach fully parabolic profiles when $l \rightarrow 2R$.
Now that for these two extremes the group speed curves are both of the classical
    $-/+$ type,
   one expects that the frequency dependence of $\vgr$ will not
   be qualitatively different
   {when $l/2R$ varies between $0$ and $1$}.
Figure~\ref{fig_vgr_para} shows that this is indeed the case: {for a number of arbitrarily chosen $l/2R$ and $\rhoi/\rhoe$},
   $\vgr$ always first sharply decreases with $\omega$ before
   increasing towards $\vai$.
{Furthermore, one sees that the group speed curves in the portion before reaching the local minima (where $\omega =\omgmin$)
   become increasingly steep when $\rhoi/\rhoe$ increases.
This happens because $\omgmin$ gets increasingly close to the angular frequency ($\omega_{\rm c}$) at the cutoff wavenumber.
Take $l/2R = 0.25$ for instance.
It turns out that when $\rhoi/\rhoe=3$ ($10$), $\omgmin = 3.82$ ($2.75$)
    and $\omega_{\rm c} = 2.76$ ($2.35$) (here and hereafter in units of $\vai/R$).
For comparison,  $\omgmin$ reads $2.27$ when $\rhoi/\rhoe=100$, in which case $\omega_{\rm c}$
    is found to be $2.24$.

The group speed behavior is further examined in}
   Figure~\ref{fig_vgr_para_contour}
   where (a) the minimum group speed ($\vgmin$) and (b) the angular frequency at which
   this minimum is attained ($\omgmin$) are shown by equally spaced contours
   in the space spanned by $l/2R$ and $\rho_{\rm i}/\rho_{\rm e}$.
One sees from Figure~\ref{fig_vgr_para_contour}a that at a given density contrast,
   $\vgmin$ tends to increase with $l/2R$, while it decreases with $\rho_{\rm i}/\rho_{\rm e}$
   when $l/2R$ is fixed.
Examining Figure~\ref{fig_vgr_para_contour}b, one sees that while $\omgmin$ tends to decrease with $\rho_{\rm i}/\rho_{\rm e}$
   at a fixed $l/2R$, its dependence on $l/2R$ is different at different values of $\rho_{\rm i}/\rho_{\rm e}$.
For $\rho_{\rm i}/\rho_{\rm e} \lesssim 4$, $\omgmin$ hardly varies with $l/2R$.
However, for larger density contrasts, $\omgmin$ tends to decrease with $l/2R$, with the tendency
   becoming increasingly prominent when $\rho_{\rm i}/\rho_{\rm e}$ increases.
{We note by passing that for typical active region loops, the density contrast tends to lie between
   $2$ and $10$~\citep[e.g.,][]{2004ApJ...600..458A}.
   Furthermore, density contrasts as large as $1000$ are not unrealistic but have been observationally deduced
       for, say, prominences~\citep[e.g.,][]{2002SoPh..208..253P, 2010SSRv..151..243L}.
   }

How about linear {TL} profiles?
Figure~\ref{fig_vgr_linear_lowden} displays the $\omega$ dependence of $\vgr$ for a number of $l/2R$ as given
    by the curves in different colors.
Here a rather modest density ratio of $3$ is chosen.
Comparing the blue and {black curves corresponding to $l/2R=0.25$ and $0.5$}, respectively,
    one sees that $\omgmin$, the angular frequency
    where the group speed attains its minimum,
    increases when the density profile becomes more diffuse.
In fact, {when $l/2R = 0.95$} (the red curve),
    $\omgmin$ moves out to infinity, and hence the ``$-$'' label.
We note that for coronal slabs with transverse density distributions describable by the generalized Epstein profile,
    \citet[][Figure~2]{1988A&A...192..343E} and \citet{1995SoPh..159..399N} showed that
    the group speed curves also experience a transition from the ``$-/+$'' to the ``$-$'' type
    when the density profile becomes less steep than the symmetric Epstein one.
What is interesting here is that, the existence of this transition depends not only on
    the profile steepness (or equivalently, the transverse lengthscale),
    but also on how the density distribution is described.
When parabolic {TL} profiles are adopted, Figure~\ref{fig_vgr_para_contour} indicates that
    the group speed curves are exclusively of the ``$-/+$'' type.
{It is necessary to point out that, when $l/2R = 0.75$,
    the group speed curve in yellow is labeled with a ``$-$'' symbol
    despite that it does attain a minimum.
The reason is that, this minimum is smaller than $\vai$ by no more than $1\%$
    and is attained at some rather large frequency.
In fact, $\omgmin$ is as large as $\sim 25 \vai/R$ in this case.
As will be demonstrated by further time-dependent computations (see Figure~\ref{fig_wavelet_lin_03_1.5}),
    the temporal evolution of impulsively generated sausage waves pertinent to this combination of parameters
    closely resembles the case where a local minimum is absent.
    }

More complications arise for higher density contrasts,
    as shown by Figure~\ref{fig_vgr_linear_highden} where $\rhoi/\rhoe = 10$.
In addition to the ``$-/+$'' behavior (see e.g., the case where $l/2R = 0.25$),
    $\vgr$ {can also} behave {in a ``$-/+/-/+$'' ($l/2R=0.75$)
    or a ``$-/+/-$'' ($l/2R=0.95$) manner.}
Evidently, these ``$-/+/-/+$'' and ``$-/+/-$'' cases are
    characterized by the existence of a local maximum $\vgmax$
    attained at some $\omgmax$, marked by the asterisks.
Note that the case labeled ``$-/+/-/+$''
    {possesses a second local minimum whose value lies between $\vai$ and the first minimum.
In what follows, by ``minimum'' we always refer to the first local minimum wherever applicable.}

Figure~\ref{fig_vgr_linear_contour} displays the distribution
    in the $l/2R-\rhoi/\rhoe$ space of
    some parameters characterizing the group speed curves.
The left (right) column concerns the first minimum (maximum).
The red and blue curves represent the boundaries separating
    different types of group speed curves, and roughly divide
    this parameter space into three regions.
The ones marked I, II, and III correspond to the cases where the group speed curves
    possess no extremum, one minimum, and more than one extrema.
{Note that in region I, for combinations of $[l/2R, \rhoi/\rhoe]$ adjacent to
    the left border, the group speed curves look similar to the yellow curve in Figure~\ref{fig_vgr_linear_lowden},
    meaning that a local minimum does exist.
However, in these cases the minima are exclusively less than $\vai$ by no more than $2\%$,
    and further time-dependent computations indicate that the corresponding impulsively generated waves
    behave as if the minima were absent.
    }

Examine the left column first.
Figure~\ref{fig_vgr_linear_contour}a suggests that $\vgmin$ increases
    with $l/2R$ but decreases with $\rhoi/\rhoe$.
On the other hand, from Figure~\ref{fig_vgr_linear_contour}b one sees that
    for low density contrasts with $\rho_{\rm i}/\rho_{\rm e} \lesssim 6$,
    $\omgmin$ increases with $l/2R$ and a tiny increase in $l/2R$
    tends to lead to an extremely rapid increase in $\omgmin$ when the blue curve is approached
    from left (note the extremely closely packed contours there).
In addition, one sees that in the region labeled III,
    $\omgmin$ tends to decrease with $l/2R$, even though this tendency is discernible
    only when $\rho_{\rm i}/\rho_{\rm e} \gtrsim 20$.
Now consider the right column.
One sees from Figure~\ref{fig_vgr_linear_contour}c that $\vgmax$
    tends to increase with $l/2R$ but shows little dependence on $\rhoi/\rhoe$.
Furthermore, Figure~\ref{fig_vgr_linear_contour}d indicates that
    $\omgmax$ tends to decrease with $l/2R$ but increase with $\rhoi/\rhoe$.


\subsection{Temporal Evolution of Impulsively Generated Sausage Waves}
The analytical and numerical results presented in the previous sections suggest that
    the behavior of the group speed curves {is} qualitatively different with
    different prescriptions of the density distribution.
However, one may question whether these characteristics can indeed be reflected
    in the temporal and wavelet signatures of impulsively generated waves.
Before examining this, let us bear in mind that the temporal evolution of impulsively generated waves
    is sensitive not only to behavior of the group speed curves,
    but also relies on
    the details of the imposed initial perturbation~\citep{2015ApJ...806...56O,2015ApJ...814..135S}.
Let us further recall that according to \citet{1986NASCP2449..347E},
    one way for characterizing the $\omega-\vgr$ curve is to distinguish between
    the S and M portions, for which a horizontal line representing a constant $\omega$
    intersects the curve at one point (multiple points).
For the classical ``$-/+$'' type, portion M
    is relevant when $h/\vai < t < h/\vgmin$, the superposition
    of multiple wavepackets therein making the signal stronger than in the earlier stage.
{Actually the ``$-/+/-/+$'' case is similar in the sense that the M portion also
    corresponds to $h/\vai < t < h/\vgmin$, the only difference being that here $\vgmin$ should
    be interpreted as the first local minimum (e.g., the yellow curve in Figure~\ref{fig_vgr_linear_highden}).
}
{Let us consider} what happens for the ``$-/+/-$'' type, for which
    $\vgmin > \vai$ (e.g., the red curve in Figure~\ref{fig_vgr_linear_highden}).
Now that the onset of portion M corresponds to a group speed of $\vgmax$,
    one expects that $\vgmax$ plays a role in regulating the amplitude modulation.
Likewise, for the monotical ``$-$'' type (e.g., the red curve in Figure~\ref{fig_vgr_linear_lowden}),
    one expects that the decay phase will not occur given the
    absence of a local extremum.
In this regard, one way to bring out the influence of different types of group speed curves
    is to examine whether the corresponding characteristic speeds can be discerned in the
    impulsively generated wave trains.

We now examine how pressureless coronal tubes respond to an impulsive, localized, axisymmetric source.
To this end, we developed a simple finite-difference code, second order accurate in both space and time,
    to solve the aximmetrical version of linearized ideal MHD equations
    in the $r-z$ plane.
The computational domain extends from $0$ to $r_{\rm M}$ in the radial (transverse) direction,
    and from $-L/2$ to $L/2$ in the axial (longitudinal) direction.
The boundaries $r=r_{\rm M}$ and $z = \pm L/2$ are placed sufficiently far such that they are irrelevant in
    determining the perturbations.
On the other hand, the boundary condition at the tube axis $r=0$
    is specified in accordance with the parity of sausage modes.
We specify the equilibrium density profile according to Equation~(\ref{eq_rho_profile}).
To initiate our computations, an initial perturbation is applied only to the transverse velocity around the origin,
   namely,
\begin{eqnarray}
 \delta v_{r}(r,z; t=0) = \frac{r}{\sigma_r}\exp\left[\frac{1}{2}\left(1-\frac{r^2}{\sigma_r^2}\right)\right]
       \exp\left(-\frac{z^2}{2\sigma_z^2}\right) .
 \label{eq_init_pert}
\end{eqnarray}
{This initial perturbation is similar in form to what was adopted in \citet{2015ApJ...814..135S},
     and ensures the parity of the generated wave trains by not displacing the loop axis.
Furthermore, a constant $\exp(1/2)$ is introduced such that the right-hand side attains a maximum of unity.
}
Here $\sigma_r$ and $\sigma_z$ determine the extent to which the initial perturbation spans in
    the transverse and longitudinal directions, respectively.
{To make sure that primarily the lowest-order modes are excited,
    $\sigma_r$ and $\sigma_z$ are both chosen to equal $R$, with the corresponding spatial profile
    of the initial perturbation shown in Figure~\ref{fig_init_pert}.}
The details of our numerical code will be presented elsewhere, here it suffices to say that
    it has been validated via an extensive set of test problems, including computations of
    both standing and propagating modes.
In addition, adopting a finer grid does not introduce any discernible difference
    to our computational results.

Let us start with an examination of the {linear transition layer} profile, and
    choose a combination $[\rho_{\rm i}/\rho_{\rm e}, l]$ of $[3, 0.5R]$
    such that the group speed curve belongs to the ``$-/+$'' type.
Figure~\ref{fig_wavelet_lin_03_0.5} displays (a) the temporal evolution
    and (b) the corresponding Morlet spectrum of
    the density perturbation $\delta\rho$ sampled at a distance $h=100R$
    along the tube axis.
This spectrum is created by using the standard wavelet toolkit devised by
    \citet{1998BAMS...79...61T}.
Note that the vertical axis in Figure~\ref{fig_wavelet_lin_03_0.5}b corresponds
    to angular frequency, and the black solid curve represents the cone of influence.
In addition, the dashed contour represents the $95\%$ confidence level,
    {computed by assuming a white-noise process for calculating the mean background spectrum
    \citep[see section 4 in][]{1998BAMS...79...61T}.}
The dotted vertical lines correspond to the arrival times of wavepackets
    traveling at the internal ($\vai$) and external ($\vae$) Alfv\'en speeds
    as well as the local minimum $\vgmin$ (see the blue curve in Figure~\ref{fig_vgr_linear_lowden}).
In agreement with the reasoning by \citet{1986NASCP2449..347E},
    one sees that the onset of the most significant power almost coincides with $h/\vai$, which marks
    the start of the M portion of the group speed curve.
Furthermore, the interval with strong power almost ends at $h/\vgmin$, beyond which
    the signal evolves into the decay phase as expected.
{One can also see that the Morlet power for $h/\vae < t < h/\vai$ shows only a rather insignificant increase in
    frequency with time, the reason being that in this time interval the arriving wavepackets
    (with group speeds between $\vai$ and $\vae$) correspond to a rather narrow frequency range.
When $h/\vai < t < h/\vgmin$, the tendency for the Morlet power to increase with time
    is more obvious due to the arrival and superposition of wavepackets
    of higher frequencies (see the blue curve in Figure~\ref{fig_vgr_linear_lowden}).
In fact, the frequency modulation in the Morlet power spectra to be presented
    can all be understood from the behavior of the corresponding group speed curves.
    }

{Still examining a linear transition layer profile with $\rho_{\rm i}/\rho_{\rm e} = 3$,
    Figure~\ref{fig_wavelet_lin_03_1.9} shows what happens when the ``$-$'' type arises by
    choosing $l$ to be $1.9R$.}
In this case, only $\vai$ and $\vae$ are relevant.
One sees that $h/\vai$ marks the end of virtually any Morelet power, in agreement with
    the expectation that the decay phase will not be present due to the
    absence of a local minimum in the corresponding group speed curve.
However, one may have noticed that for the parabolic {transition layer} profile with the same parameters,
    the group speed curve is not remarkably different in that the minimum $\vgmin$ is not too far
    from $\vai$ (actually $\vgmin = 0.967\vai$, see the red curve in Figure~\ref{fig_vgr_para}a).
The dispersion pertinent to $\vgmin < \vgr < \vai$ is not significant, and
    one may see the group speed curve as being monotonically decreasing,
    as is the case for the relevant linear profile.
One may then ask whether it is possible to discern this subtle difference between the linear and parabolic profiles.
To address this, Figure~\ref{fig_wavelet_par_03_1.9} examines the pertinent parabolic case,
    from which one sees that $h/\vgmin$ indeed shows up by
    marking the end of the most significant Morlet power.
In addition, now the decay phase is present due to the existence of $\vgmin$, despite that it is
    indeed very close to $\vai$.
{Another case that needs to be contrasted here is the linear transition layer profile
    with $\rho_{\rm i}/\rho_{\rm e} = 3$ and $l=1.5 R$.
As shown by Figure~\ref{fig_wavelet_lin_03_1.5}, one sees that in this case the temporal evolution
    of the Morlet power is similar to what happens when $l=1.9R$.
In particular, even though in this case a local minimum does exist as shown by the yellow curve in Figure~\ref{fig_vgr_linear_lowden},
    the decay phase is so weak that it does not show up in the Morlet power (see the part for $t>h/\vgmin$
    in Figure~\ref{fig_wavelet_lin_03_1.5}b).
Actually this behavior is typical of the temporal signatures of impulsively generated waves for linear transition layer profiles
    with combinations of $[\rhoi/\rhoe, l]$ not far the left border in region I of Figure~\ref{fig_vgr_linear_contour}.
And that is why we place these combinations in region I, in which the group speed curves, strictly speaking,
    should be monotonical and possess no extremum.

The question now is what causes the difference between Figures~\ref{fig_wavelet_par_03_1.9} and \ref{fig_wavelet_lin_03_1.5}?
The corresponding group speed curves indicate that while the local minima are both close to $\vai$, the specific locations of
    the dimensionless angular frequencies $\omgmin R/\vai$ where these minima are attained are different by almost an order of magnitude.
As a result, the dimensionless wavenumbers $kR$ pertinent to these frequencies are substantially different.
To be specific, this $kR$ reads $2.56$ for the parabolic TL profile with $[\rhoi/\rhoe, l] = [3, 1.9 R]$,
    whereas it reads $24.2$ for the linear TL profile with $[\rhoi/\rhoe, l] = [3, 1.5 R]$.
Given that the initial perturbation as given by Equation~(\ref{eq_init_pert}) primarily excites wavepackets with longitudinal
    wavenumbers not too large relative to $1/\sigma_z = 1/R$,
    the wavepackets with $\vgmin < \vgr <\vai$ pertinent to Figure~\ref{fig_wavelet_par_03_1.9} are substantially stronger
    than those pertinent to Figure~\ref{fig_wavelet_lin_03_1.5}.
Consequently, the Morlet power for $t>h/\vai$ in Figure~\ref{fig_wavelet_lin_03_1.5} is negligible whereas it is not so
    in Figure~\ref{fig_wavelet_par_03_1.9}.
In this sense, the difference between the two figures highlights the importance of the description of the initial perturbation
    in determining the temporal signatures of impulsively generated sausage waves, as was pointed out
    by \citet{2015ApJ...806...56O} and \citet{2015ApJ...814..135S}.
    }

{Now turn to a linear transition layer profile with $\rhoi/\rhoe = 10$ and $l=1.5 R$, for which the
   group speed curve is labeled ``$-/+/-/+$'' in Figure~\ref{fig_vgr_linear_highden}.
The temporal and spectral signatures of the corresponding impulsively generated wave trains
   are given by Figure~\ref{fig_wavelet_lin_10_1.5}, where one can see that these signatures
   are similar to what happens in the ``$-/+$'' case in the sense that $h/\vgmin$ marks
   the end of the most significant Morlet power.
The reason for this similarity is that the M portion in the group speed curve corresponds to $\vgmin < \vgr < \vai$,
   the extra maximum and minimum being irrelevant in this case because they lie between $\vgmin$ and $\vai$.
}

How about the ``$-/+/-$'' type?
To show this, let us examine a linear {transition layer} profile
    with $\rho_{\rm i}/\rho_{\rm e} = 10$ and $l = 1.9 R$.
The corresponding group speed diagram (the red curve in Figure~\ref{fig_vgr_linear_highden})
    indicates that the M portion starts where the group speed equals $\vgmax$.
From the temporal evolution and Morelet spectrum as presented in Figure~\ref{fig_wavelet_lin_10_1.9},
    one sees that indeed $h/\vgmax$ characterizes the onset of
    the most significant power.
We note that higher-order sausage modes are likely to have been excited for the interval $h/\vgmax < t < h/\vai$,
    but they cannot account for the increase in the most significant part of the Morlet power.
The reason is, the critical wavenumber for the second branch is $1.48/R$, resulting in an angular frequency
    of $\omega = 4.68 \vai/R$.
However, the majority of the Morlet power corresponds to some $\omega\lesssim 4 \vai/R$, meaning that
    the increase in Morelet power derives from the superposition of wavepackets
    with $\vgr < \vgmax$ along the group speed curve for the lowest-order mode.
Now move on to the vertical line representing $h/\vai$.
One sees that it marks the start of the decay phase, and consequently
    the Morelet power starts to decrease at this point.

\section{SUMMARY}
\label{sec_conc}
Quasi-periodic propagating disturbances have been seen in a substantially number of
    coronal structures.
While intuitively speaking this quasi-periodicity has to do with quasi-periodicities
    in the wave sources, it is equally possible to be caused by an impulsive driver.
In this latter scenario, the frequency dependence of the longitudinal group speed is crucial for determining the
    temporal and spectral signatures of impulsively generated disturbances.
On the other hand, while top-hat transverse density profiles have been extensively examined,
    little is known on how other density distributions may impact the group speed diagrams.
To address this issue, we focused on sausage modes throughout, and
    started with an analytical study on the fully parabolic profile to show that
    the group speed diagrams are qualitatively the same as in the case of top-hat profiles.
Both show the classical ``$-/+$'' dependence on angular frequency $\omega$.
We then contrasted two more general profiles, labeled ``parabolic'' and ``linear'', by capitalizing
    on our previous theoretical study on sausage waves in coronal tubes with transverse density profiles
    that comprise a uniform cord, a uniform external medium, and a transition layer (TL) connecting the two.
Here two parameters are relevant, one being the density contrast $\rhoi/\rhoe$ between coronal tubes and their surroundings,
    and the other being the dimensionless TL width $l/R$ where $R$ is the mean tube radius.

For parabolic profiles, the group speed $\vgr$ shows the typical ``$-/+$'' behavior
    regardless of $\rhoi/\rhoe$ and $l/R$.
However, for linear profiles, $\vgr$ shows a much richer variety of $\omega$ dependence.
For low density contrasts with $\rhoi/\rhoe \lesssim 6$, the $\omega-\vgr$ curves
    experience a transition from the ``$-/+$'' to the ``$-$'' type when $l/R$ exceeds
    some critical value.
On the other hand, for higher density contrasts, the ``$-/+/-/+$'' and ``$-/+/-$'' types
    arise, meaning that the $\omega-\vgr$ curves can posses a local maximum $\vgmax$
    in addition to a minimum $\vgmin$.
With time-dependent computations, we further showed that the different behavior of group speed curves,
    the characteristic speeds $\vgmin$ and $\vgmax$ in particular,
    is indeed reflected in the temporal evolution and Morlet spectra
    of impulsively generated wave trains.

\begin{figure}
\centering
\includegraphics[width=0.6\columnwidth]{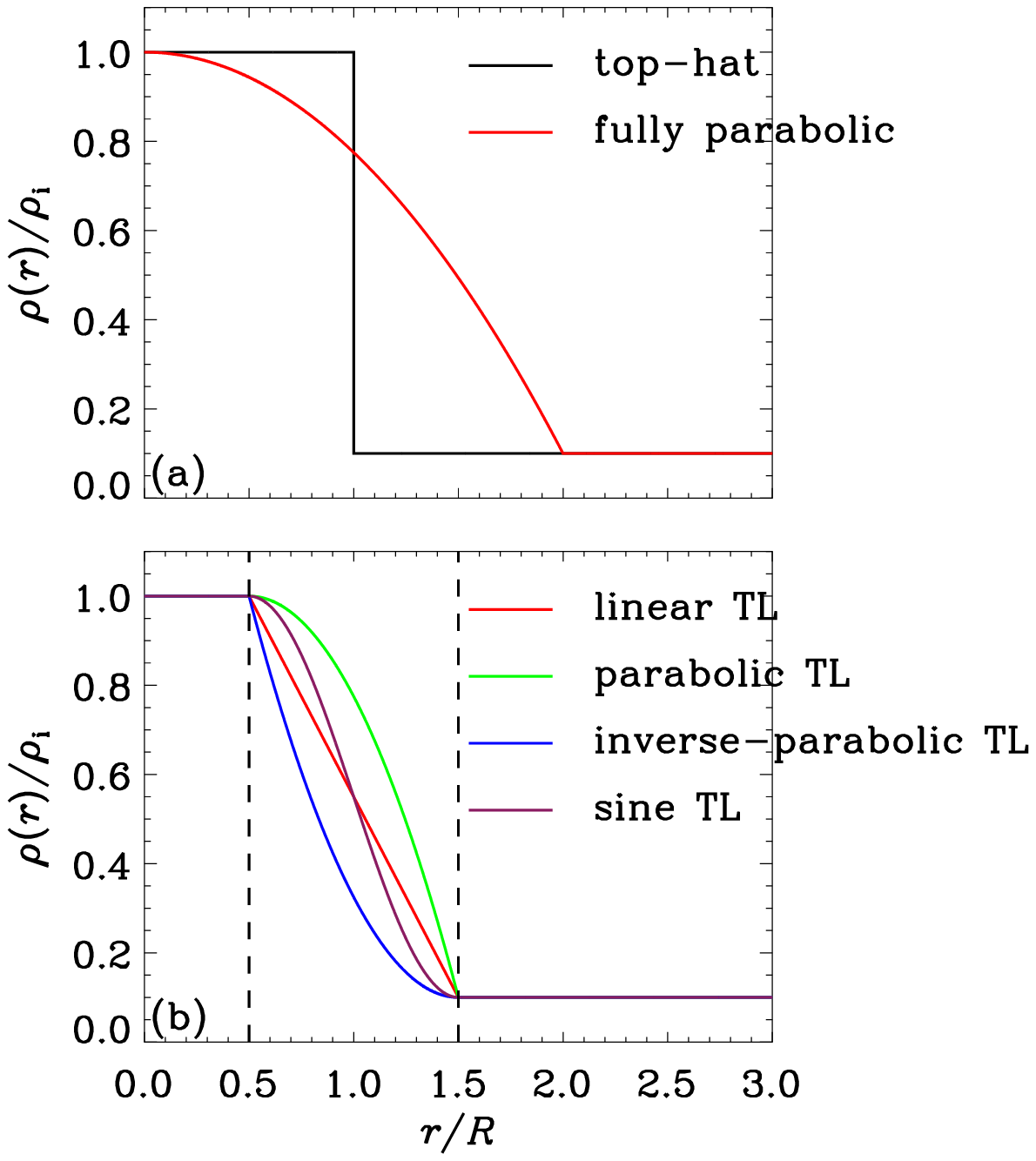}
\caption{
 Transverse density profiles examined in this study as a function of $r$.
 Top-hat and fully parabolic profiles are displayed in panel (a).
 In panel (b), the profiles differ only in a transition layer (TL)
     sandwiched between the internal (with a uniform density $\rho_{\rm i}$)
     and external (with a uniform density $\rho_{\rm e}$) portions.
 The TL is located between
     $R-l/2$ and $R+l/2$, with $R$ being the mean tube radius.
 For illustration purposes, the density contrast $\rho_{\rm i}/\rho_{\rm e}$ is chosen to be $10$ for both panels,
     while $l$ is chosen to be $R$ in panel (b).
}
 \label{fig_illus_profile}
\end{figure}

\begin{figure}
\centering
\includegraphics[width=0.8\columnwidth]{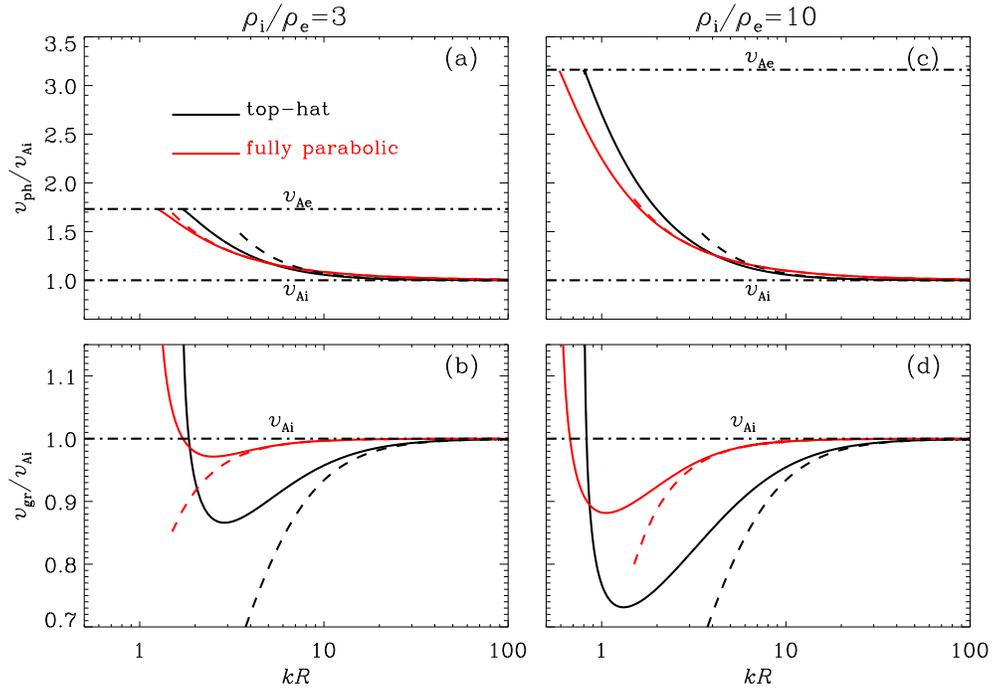}
 \caption{
 Dependence on axial wavenumber of the {phase (the upper row) and  group (lower) speeds
     for top-hat (the black curves) and fully parabolic (red) profiles}.
 The solid curves represent numerical solutions to the relevant dispersion relations,
     and the dashed curves represent approximate expressions at large axial wavenumbers.
 For illustration purposes, {two values, $3$ (the left column) and $10$ (right),
     are chosen for the density contrast $\rho_{\rm i}/\rhoe$}.
}
 \label{fig_vphvgkdep_tophatfullypara}
\end{figure}

\begin{figure}
\centering
\includegraphics[width=0.6\columnwidth]{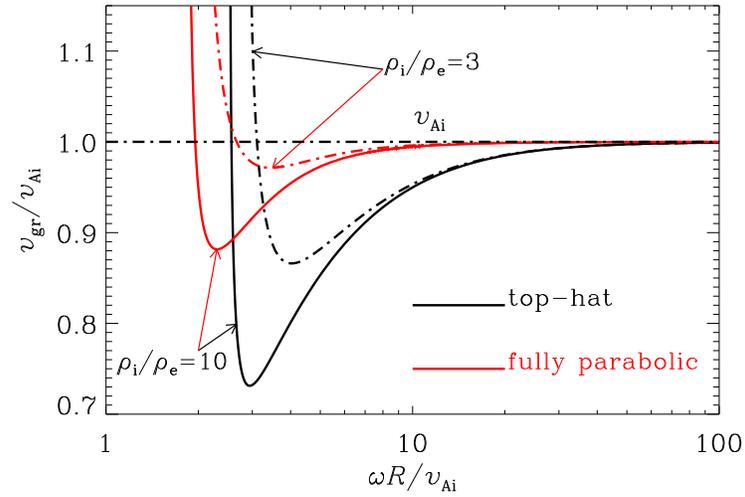}
 \caption{
 Dependence on the angular frequency of the group speeds for
  {top-hat (the black curves) and fully parabolic (red) profiles.
       Two values, $3$ (the dash-dotted curves) and $10$ (solid),
       are chosen for the density contrast $\rho_{\rm i}/\rhoe$.}
}
 \label{fig_vgr_tophatfullypara}
\end{figure}

\begin{figure}
\centering
\includegraphics[width=0.6\columnwidth]{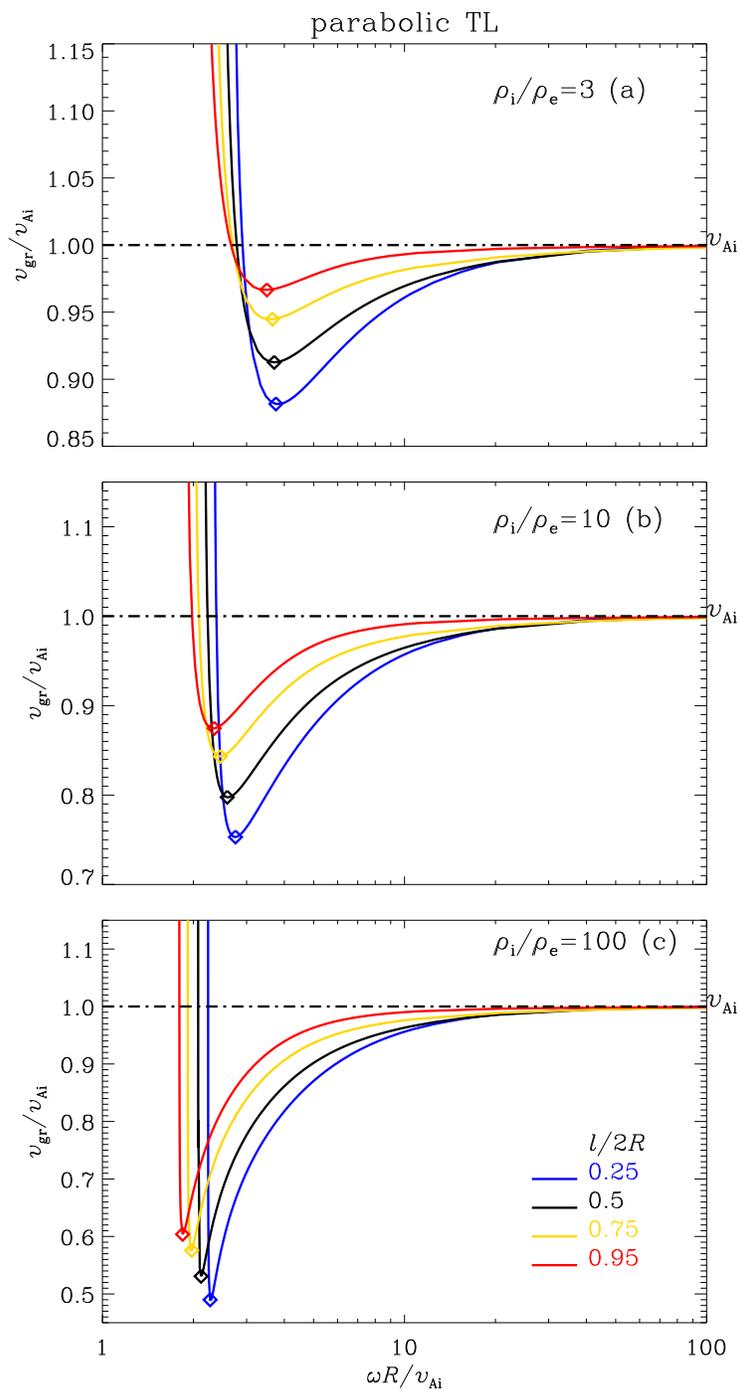}
 \caption{
 Dependence on the angular frequency of the group speeds
     for parabolic {transition layer profiles with $\rho_{\rm i}/\rhoe$ being
     (a) $3$, (b) $10$, and (c) 100}.
 A number of arbitrarily chosen $l/2R$ are examined as labeled.
 The diamonds represent the minima in the curves.
}
 \label{fig_vgr_para}
\end{figure}

\begin{figure}
\centering
\includegraphics[width=0.9\columnwidth]{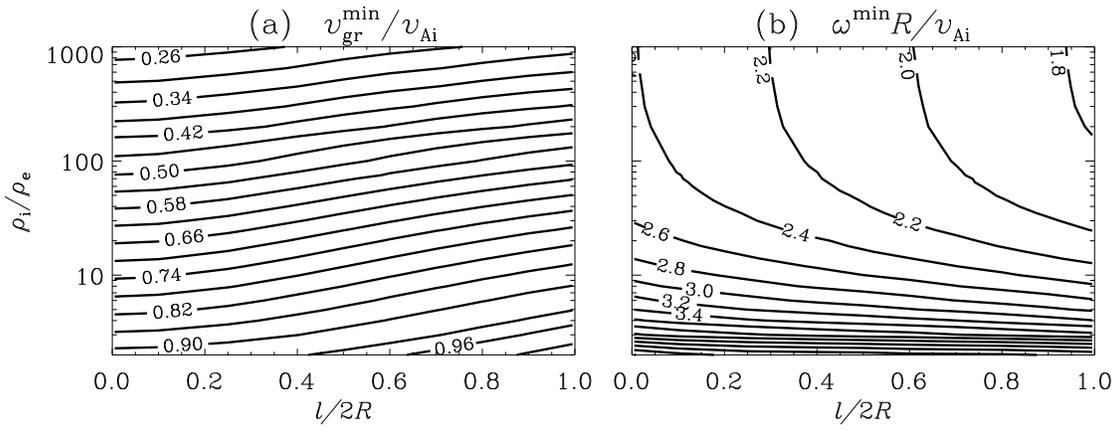}
 \caption{
 Distribution of the minimum group speed ($\vgmin$)
     and the angular frequency where $\vgmin$ is attained ($\omgmin$)
     in the parameter space spanned by {$l/2R$ and $\rhoi/\rhoe$.
 Parabolic transition layer profiles are examined here.}
}
\label{fig_vgr_para_contour}
\end{figure}

\begin{figure}
\centering
\includegraphics[width=0.6\columnwidth]{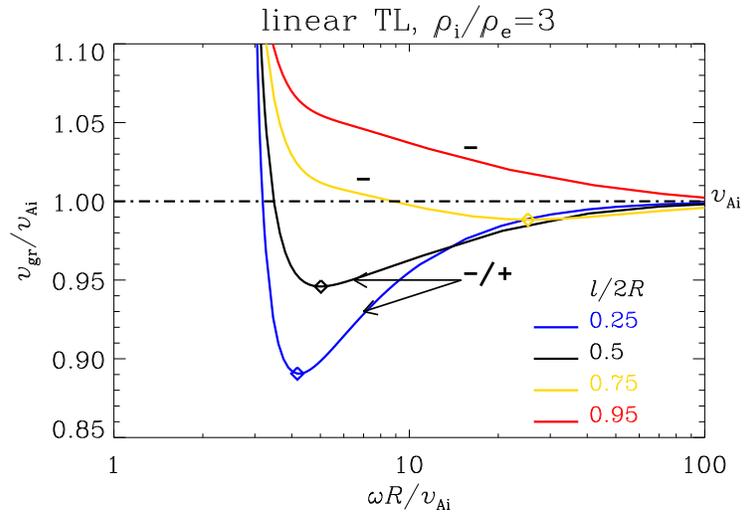}
 \caption{
 Dependence on the angular frequency of the group speeds
     for {linear transition layer} profiles with a number of different $l/2R$ as labeled.
 Here $\rho_{\rm i}/\rhoe$ is chosen to be $3$.
 The diamonds represent the minima in the curves.
 See text for the meaning of the symbols ``$-$'' and ``$+$''.
}
 \label{fig_vgr_linear_lowden}
\end{figure}

\begin{figure}
\centering
\includegraphics[width=0.6\columnwidth]{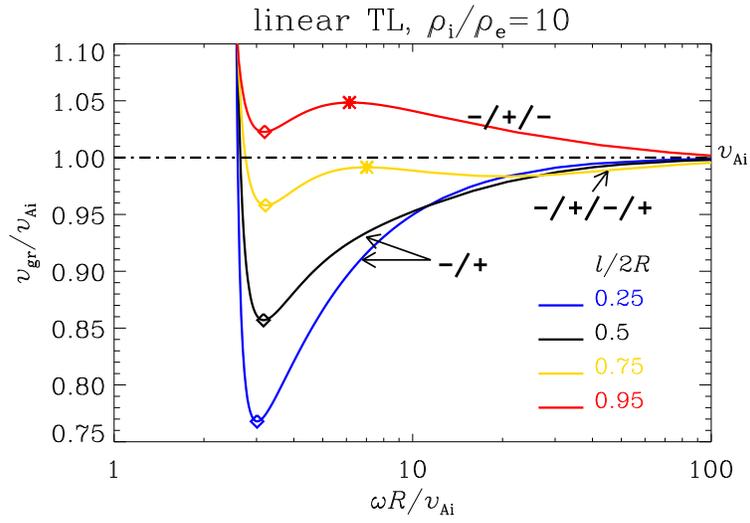}
 \caption{
 Similar to Figure~\ref{fig_vgr_linear_lowden} but for $\rho_{\rm i}/\rhoe = 10$.
 The asterisks represent the maxima in the curves.
 }
 \label{fig_vgr_linear_highden}
\end{figure}

\begin{figure}
\centering
\includegraphics[width=0.9\columnwidth]{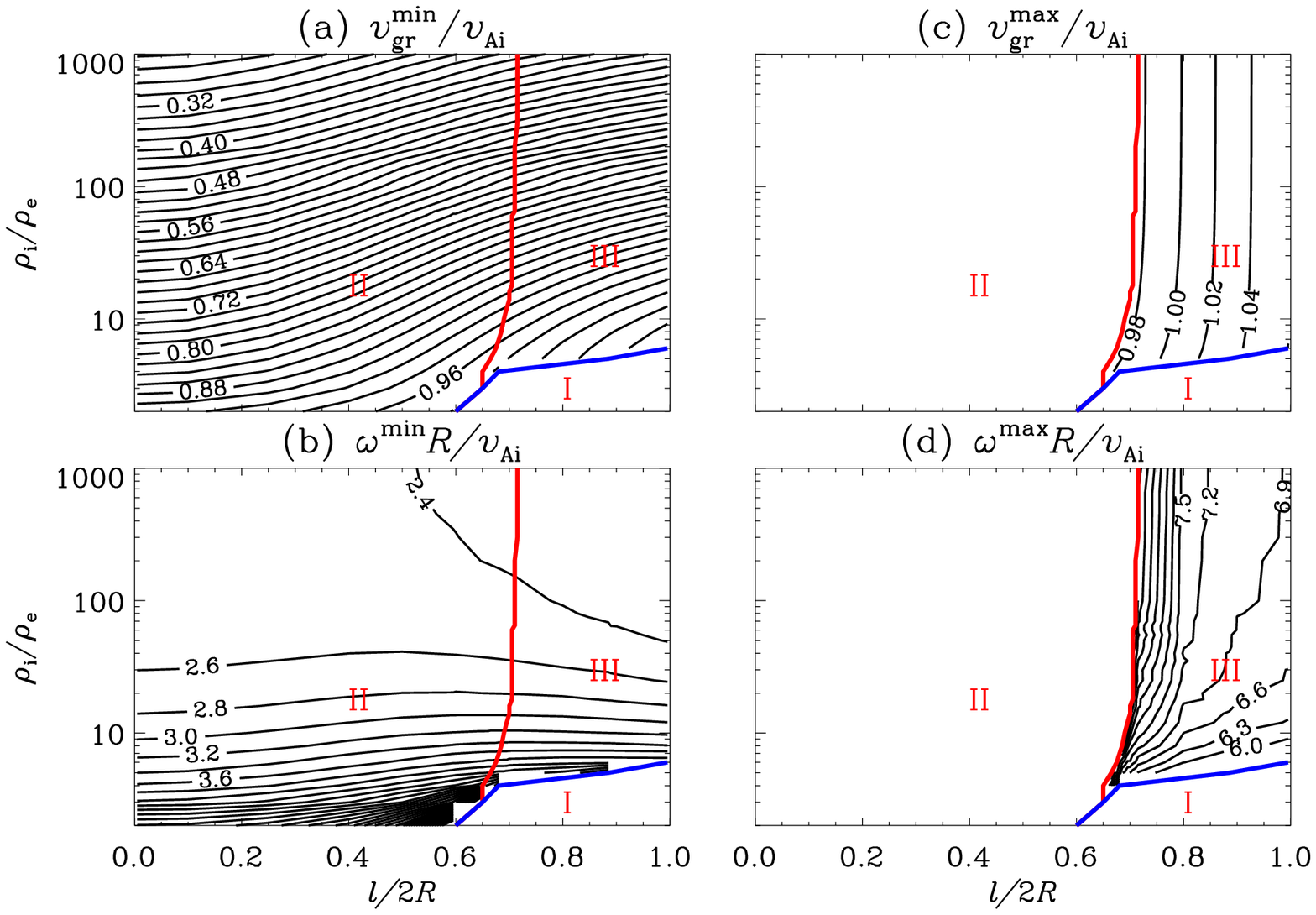}
 \caption{
 Distribution of $\vgmin$, $\omgmin$, $\vgmax$ and $\omgmax$
     in the parameter space spanned by $l/2R$ and $\rhoi/\rhoe$.
 {Linear transition layer} profiles are examined here.
 The red and blue curves represent where the group speed curve transitions from
     one type to another.
 The areas labeled I, II, and III correspond to the cases where the group speed curve possesses
     no extremum, one extremum, and more than one extrema.
 See text for details.
}
 \label{fig_vgr_linear_contour}
\end{figure}

\begin{figure}
\centering
\includegraphics[width=0.8\columnwidth]{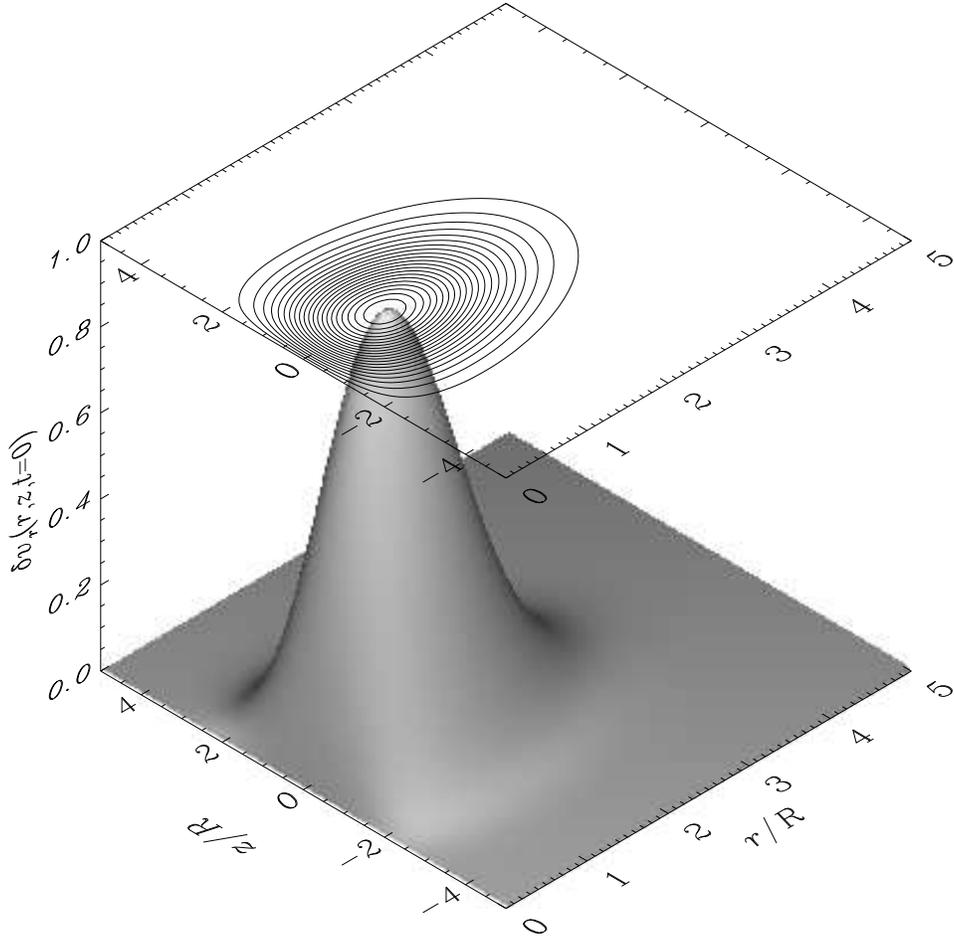}
 \caption{
 {Initial perturbation to the transverse velocity as a function of $r$ and $z$,
    shown as a contour plot superimposed on a surface plot.
 Here both $\sigma_r$ and $\sigma_z$ are chosen to be $R$ (see Equation~\ref{eq_init_pert}).
 }
 }
 \label{fig_init_pert}
\end{figure}

\begin{figure}
\centering
\includegraphics[width=0.8\columnwidth]{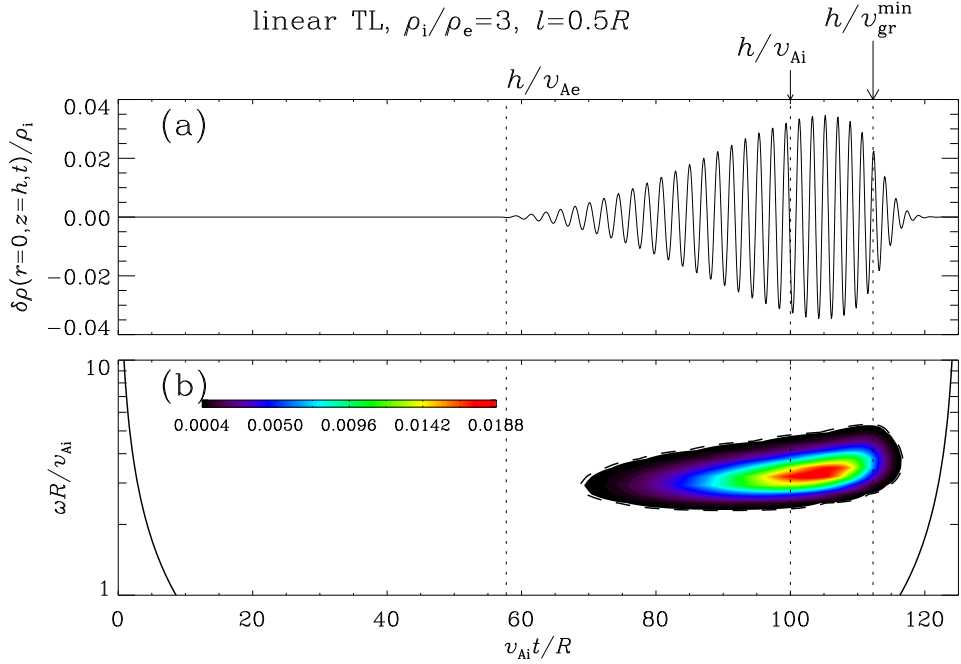}
 \caption{
 Density perturbation $\delta\rho$ at a distance $h=100R$
     from the impulsive source along the axis of a coronal tube
     with a {linear transition layer} profile.
 Here the density contrast $\rho_{\rm i}/\rho_{\rm e}$ is $3$,
     and the transverse density lengthscale $l = 0.5 R$.
 In addition to the the temporal evolution (panel a),
     the corresponding Morlet spectrum is also shown (panel b), for which
     the vertical axis corresponds to the angular frequency $\omega$.
 The black solid curve in (b) represents the cone of influence,
     and the area inside the dashed contour indicates where the Morlet power
     exceeds the $95\%$ confidence level.
 The dotted vertical lines in both panels correspond to the arrival times
     of wavepackets traveling at some characteristic group speeds.
 See text for details.
}
 \label{fig_wavelet_lin_03_0.5}
\end{figure}

\begin{figure}
\centering
\includegraphics[width=0.8\columnwidth]{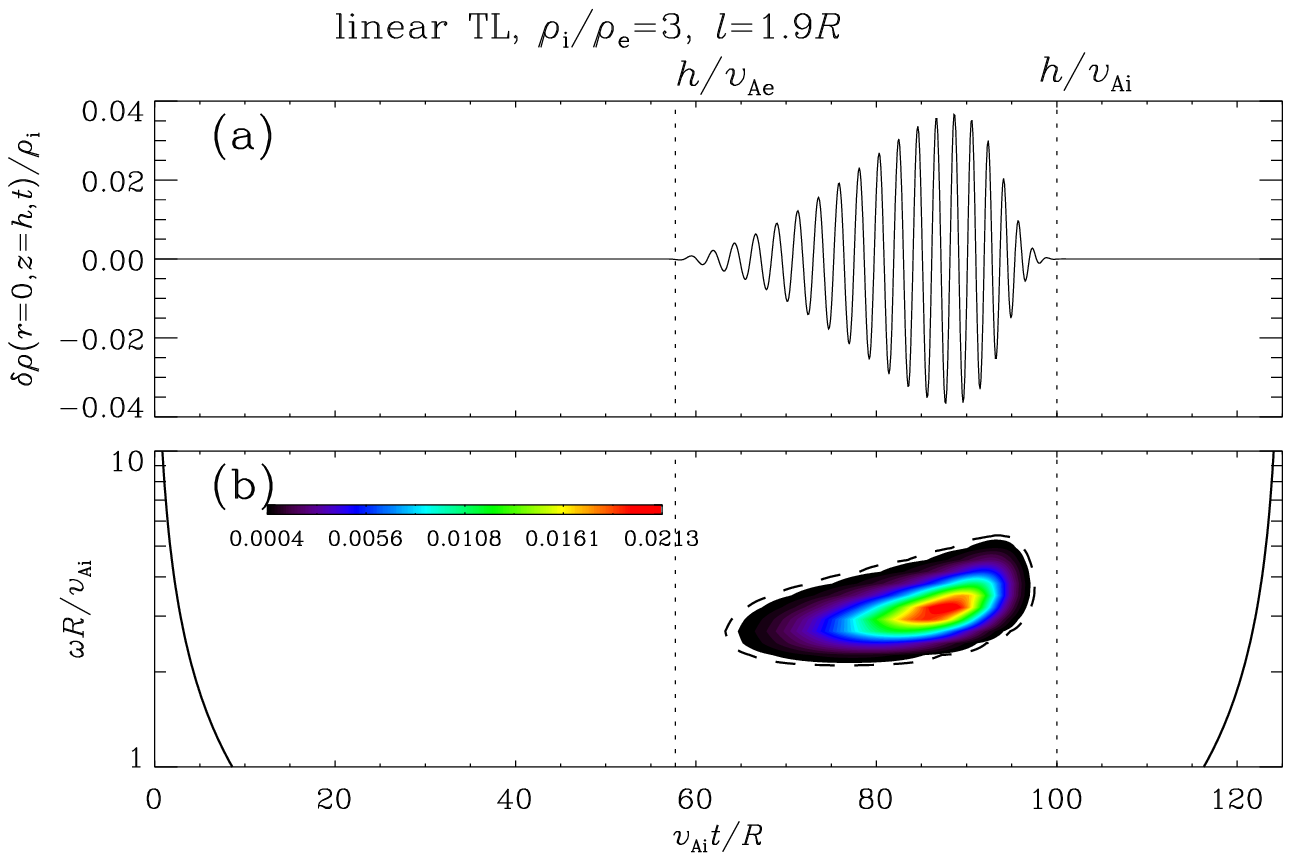}
 \caption{
 Similar to Figure~\ref{fig_wavelet_lin_03_0.5}, but for a {linear transition layer} profile
     with $\rho_{\rm i}/\rho_{\rm e} = 3$ and $l = 1.9R$.
}
 \label{fig_wavelet_lin_03_1.9}
\end{figure}

\begin{figure}
\centering
\includegraphics[width=0.8\columnwidth]{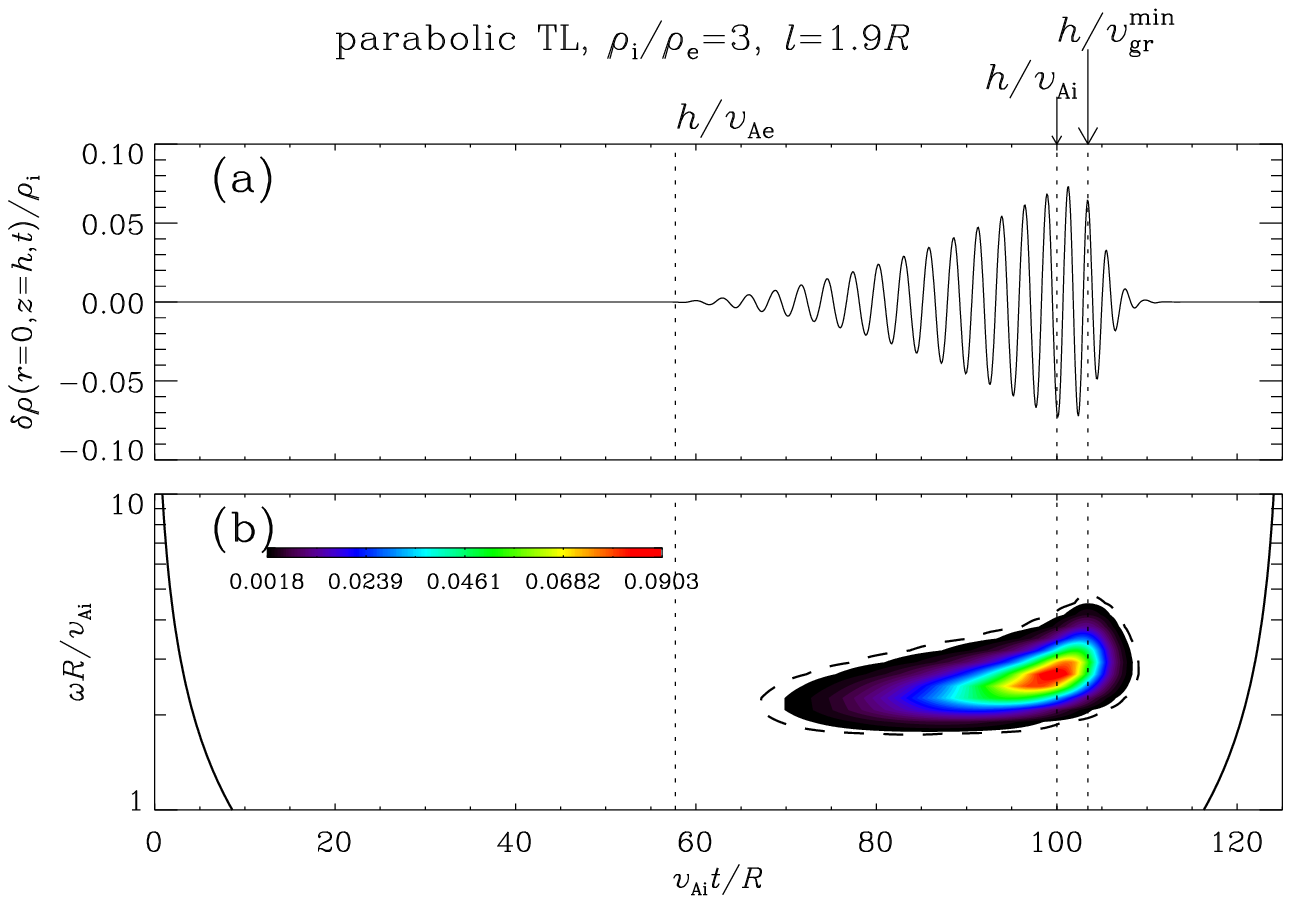}
 \caption{
 Similar to Figure~\ref{fig_wavelet_lin_03_0.5}, but for a {parabolic transition layer} profile
     with $\rho_{\rm i}/\rho_{\rm e} = 3$ and $l = 1.9R$.
}
 \label{fig_wavelet_par_03_1.9}
\end{figure}

\begin{figure}
\centering
\includegraphics[width=0.8\columnwidth]{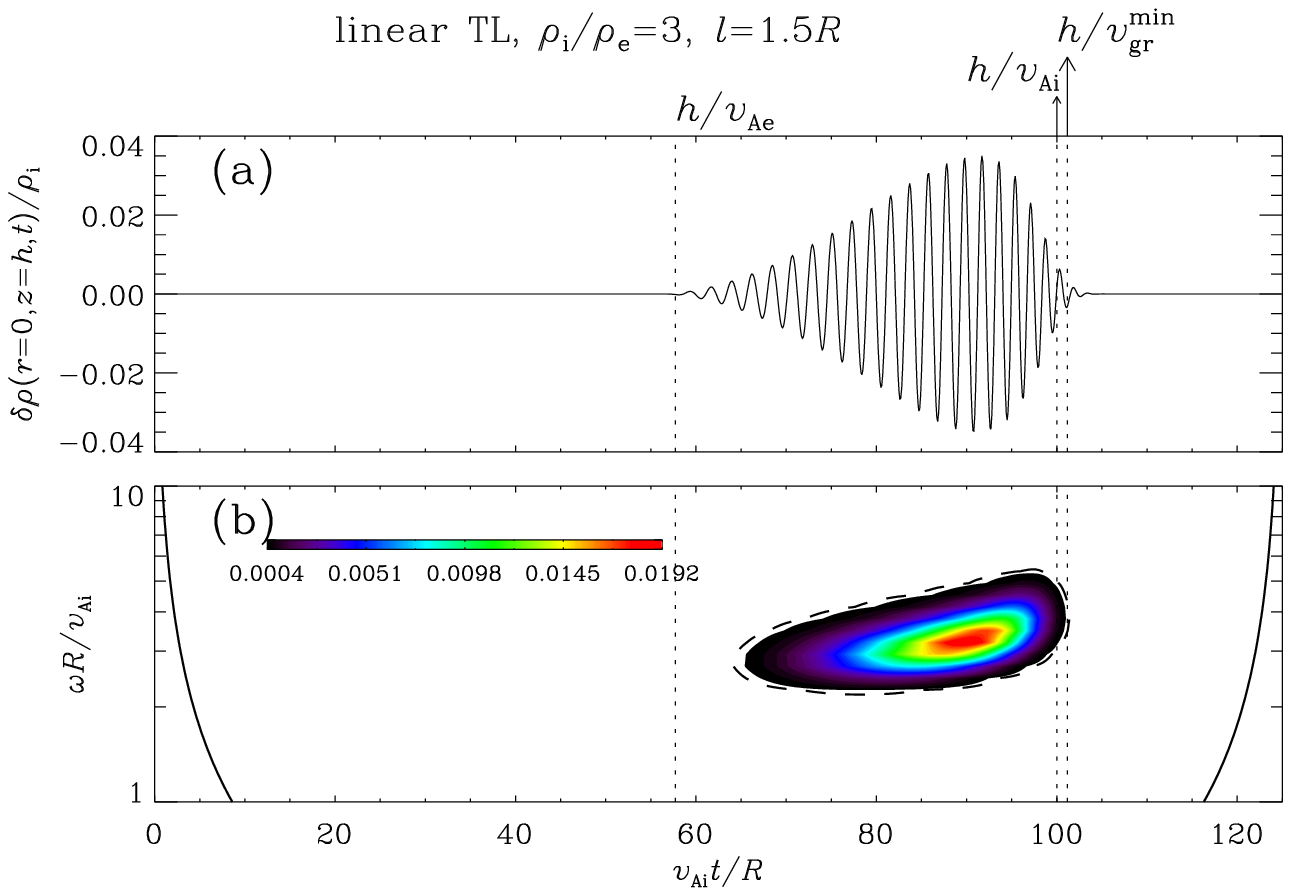}
 \caption{
 Similar to Figure~\ref{fig_wavelet_lin_03_0.5}, but for a {linear transition layer} profile
     with $\rho_{\rm i}/\rho_{\rm e} = 3$ and $l = 1.5R$.
}
 \label{fig_wavelet_lin_03_1.5}
\end{figure}

\begin{figure}
\centering
\includegraphics[width=0.8\columnwidth]{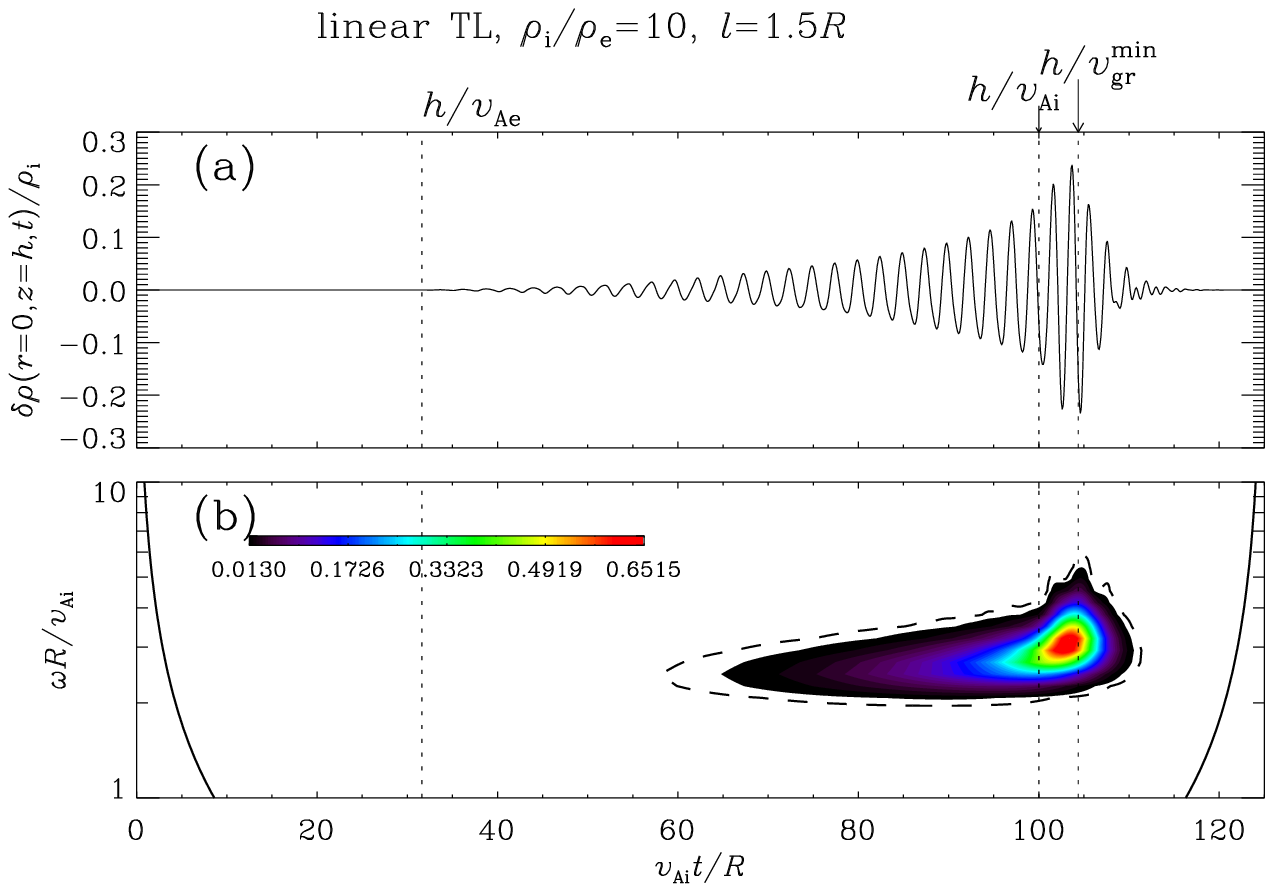}
 \caption{
 Similar to Figure~\ref{fig_wavelet_lin_03_0.5}, but for a {linear transition layer} profile
     with $\rho_{\rm i}/\rho_{\rm e} = 10$ and $l = 1.5R$.
}
 \label{fig_wavelet_lin_10_1.5}
\end{figure}

\begin{figure}
\centering
\includegraphics[width=0.8\columnwidth]{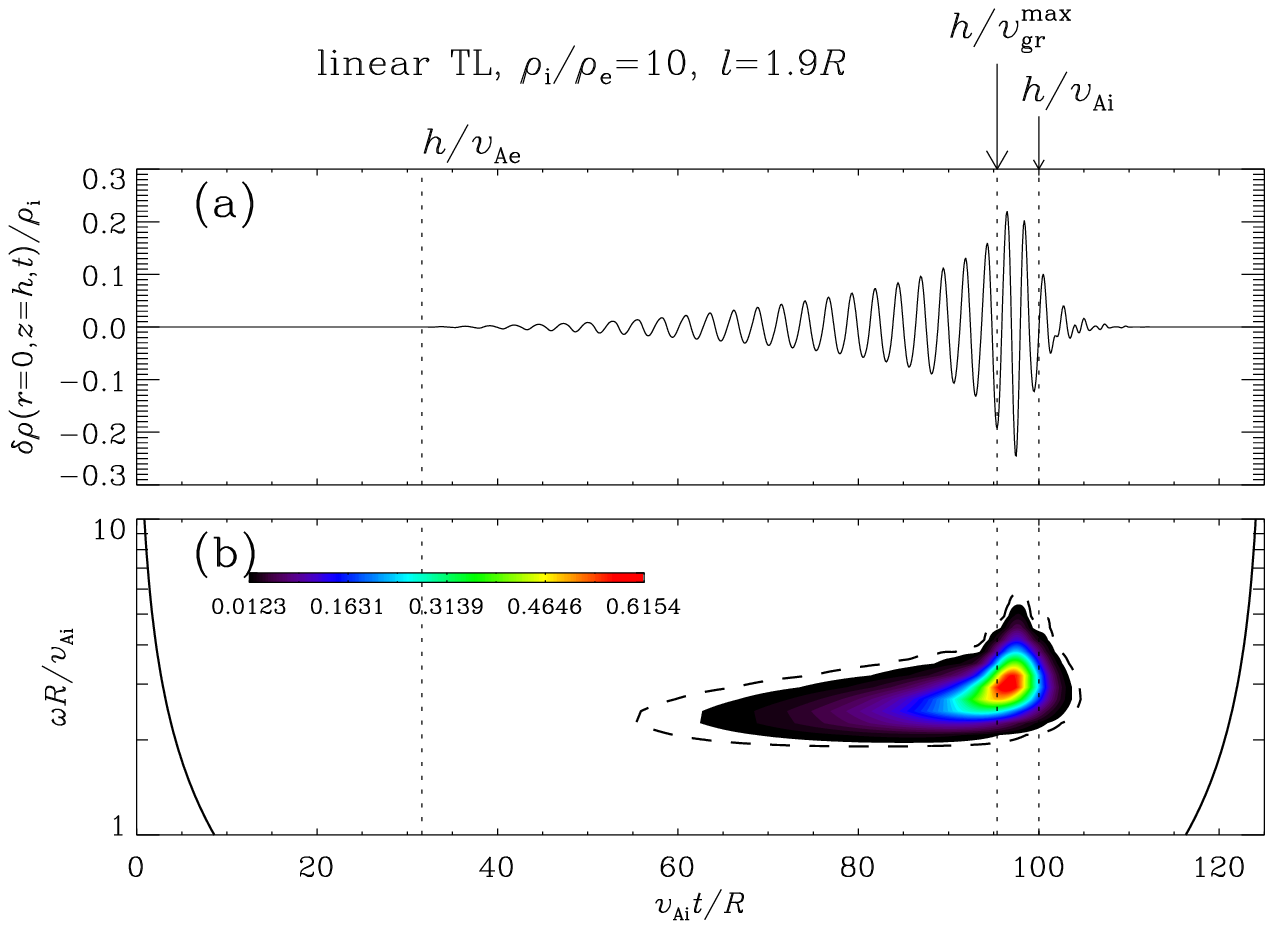}
 \caption{
 Similar to Figure~\ref{fig_wavelet_lin_03_0.5}, but for a {linear transition layer} profile
     with $\rho_{\rm i}/\rho_{\rm e} = 10$ and $l = 1.9R$.
}
 \label{fig_wavelet_lin_10_1.9}
\end{figure}

Given that the density structuring transverse to coronal loops remains largely unknown,
    it is reasonable to ask how representative the presented group speed diagrams are.
In the present study, we detailed the linear profiles and showed that
    the $\omega$ dependence other than the ``$-/+$'' behavior takes place
    in a rather extensive range.
In fact, we have also experimented with the profiles labeled ``inverse parabolic'' and ``sine'' in our Paper I
    {(illustrated in Figure~\ref{fig_illus_profile}b)},
    and found that the results are qualitatively similar to the ``linear'' and ``parabolic'' cases, respectively.
While it is admittedly impossible to exhaust the possible prescriptions for the transverse density distribution,
    our computations suggest that the group speed curves can indeed behave in a manner
    different from the classical ``$-/+$'' type.
On top of that, these characteristics can be discerned in
    the corresponding temporal and spectral evolution of impulsively generated wave trains.
Observationally, this means that these wave trains can be employed not only to probe such parameters as
    density contrasts and density profile steepness, but also to tell the form that best describes
    the transverse density distribution.

\acknowledgments
This work is supported by the 973 program 2012CB825601,
    the National Natural Science Foundation of China (BL:41174154, 41274176, and 41474149, MX:41374175),
    and by the Provincial Natural Science Foundation of Shandong via Grant JQ201212.

\bibliographystyle{apj}
\bibliography{impls}

\IfFileExists{\jobname.bbl}{} {\typeout{}
\typeout{****************************************************}
\typeout{****************************************************}
\typeout{** Please run "bibtex \jobname" to obtain} \typeout{**
the bibliography and then re-run LaTeX} \typeout{** twice to fix
the references !}
\typeout{****************************************************}
\typeout{****************************************************}
\typeout{}}

\clearpage

\clearpage

\clearpage

\clearpage

\clearpage

\clearpage

\clearpage

\clearpage

\clearpage

\clearpage

\clearpage

\clearpage

\clearpage

\clearpage

\clearpage

\end{document}